\documentclass[conference,10pt]{IEEEtran}

\usepackage[linesnumbered,ruled]{algorithm2e}
\usepackage{graphicx}
\usepackage[T1]{fontenc}
\usepackage[table,xcdraw]{xcolor}
\usepackage{multirow}
\usepackage{dblfloatfix}
\usepackage{caption}
\usepackage[numbers,sort&compress]{natbib}
\usepackage{url}
\usepackage[caption=false]{subfig}
\usepackage{enumitem}
\usepackage{amsmath}
\usepackage{breqn}
\usepackage{footnote}
\usepackage[symbol]{footmisc}

\begin{document}

\title{LEADER: Low Overhead Rank Attack Detection for Securing RPL based IoT}

\author{\IEEEauthorblockN{Somnath Karmakar}
\IEEEauthorblockA{Indian Institute of Engineering \\ Science and Technology\\
Howrah, India\\
Email: somnathkarmakar73@gmail.com}
\and
\IEEEauthorblockN{Jayasree Sengupta}
\IEEEauthorblockA{Indian Institute of Engineering \\ Science and Technology\\
Howrah, India\\
Email: jayasree202@gmail.com}
\and
\IEEEauthorblockN{Sipra Das Bit}
\IEEEauthorblockA{Indian Institute of Engineering \\ Science and Technology \\ Howrah, India \\ Email : sdasbit@yahoo.co.in}}

\maketitle
\begin{abstract}
   In recent times researchers have found several security vulnerabilities in the Routing Protocol for Low power and Lossy network (RPL), amongst which rank attack is a predominant one causing detrimental effects on the network by creating a fake topology. To address this concern, we propose a low-overhead rank attack detection scheme for non-storing mode of RPL used in IoT to deal with both increased and decreased rank attacks. Accordingly, we have modified the RPL Destination Oriented Directed Acyclic Graph (DODAG) formation algorithm to detect rank attacks during topology formation and maintenance. The distributed module of the algorithm runs in all the participating nodes whereas the centralized module runs in the sink. Unlike many existing schemes, instead of sending additional control message, we make the scheme low-overhead by simply modifying the DAO control message. Additionally, a lightweight Message Authentication Code (HMAC-LOCHA) is used to verify the integrity and authenticity of the control messages exchanged between nodes and the sink. The correctness of the proposed scheme is established through a concrete proof using multiple test case scenarios. Finally, the performance of the proposed scheme is evaluated both theoretically and through simulation in Contiki-based Cooja simulator. Theoretical evaluation proves the scheme’s energy efficiency. Simulation results show that our scheme outperforms over a state-of-the-art rank attack detection scheme in terms of detection accuracy, false positive/negative rate and energy consumption while also keeping acceptable network performance such as improved detection latency and at par packet delivery ratio.
\end{abstract}
\IEEEpeerreviewmaketitle

\section{Introduction}

With the enormous growth of the Internet and the evolution of IoT \cite{c17,c16}, conventional routing protocols can no longer accommodate the large number of nodes. Thus, Routing protocol for Low Power and Lossy Networks (RPL) \cite{c2} has been developed by Internet Engineering Task Force (IETF) especially for Low-power and Lossy Networks (LLNs) and quickly gained popularity among the research community. RPL uses a Destination Oriented Directed Acyclic Graph (DODAG) to represent the topology of the network where the nodes maintain the network state information. However, RPL doesn’t have any in-built security feature including basic securities such as integrity and authentication. Thus, it is vulnerable to several attacks even during topology formation and maintenance phase. Rank attack is one such attack which can disrupt the entire network and also cause other attacks like selective forwarding, sinkhole etc. apart from deteriorating the network performance.

To the best of our knowledge, a few works \cite{r2,r3,r4,r6,r7,r8,r1,r5} have so far been reported on dealing with rank attacks in RPL. In one such work \cite{r2}, the authors have proposed a mechanism known as TRAIL to detect decreased rank attacks and exclude individual attackers by examining the integrity of the RPL routing structure. Another work \cite{r3} has proposed an intrusion detection system to detect manipulations performed in DIO messages for launching decreased rank attacks only. Shafique et al. \cite{r4} have proposed a sink-based intrusion detection system (SBIDS) to detect decreased rank attacks in RPL. They have also used AES-128 to preserve integrity of DAO control messages. The authors in \cite{r6} have proposed a lightweight Identity-based Online/Offline signature mechanism to prevent rank spoofing by signing the rank of a node. This mechanism helps to deal with decreased rank attacks while assuring non-repudiation as well. The work \cite{r7} has proposed a highly scalable, seven layer neural network model to detect decreased rank attacks in general. Airehrour et al. \cite{r8} have proposed a trust-aware routing protocol, named SecTrust-RPL to provide protection against decreased rank attacks while optimizing performance in terms of detection and packet loss rate. Although the work \cite{r1} hasn’t dealt with rank attacks directly, it has prevented sinkhole and selective forwarding attacks in RPL by using rank manipulation information. Lastly, Sahay et al. \cite{r5} have conducted an assessment on the severity of the effects caused by different types of rank attacks in RPL.

From the above discussion it is clear that the state-of-the-art schemes are either computationally intensive (e.g. AES-128, hash chain authentication) or have high communication overhead in terms of extra control message exchange. Further, most of the schemes are devised to detect/prevent decreased rank attacks only. This motivates us to fill the research gap by putting forth the following contributions:

\begin{itemize}[leftmargin=*]
    \item We propose a low-overhead rank attack detection scheme for securing RPL during topology formation and maintenance.
    \begin{itemize}[leftmargin=*]
        \item detect both increased and decreased rank attacks.
        \item made low-overhead by simply modifying the DAO control message instead of sending any additional control message.
        \item preserve basic security in terms of integrity and authenticity of the control message exchanged by using a low-overhead MAC algorithm.
    \end{itemize}
    \item We establish the correctness of our scheme through a concrete proof using various test cases.
    \item Finally, we evaluate and establish the practicality of our scheme both theoretically and experimentally.
    \end{itemize}

The rest of the paper is structured as follows. Section II briefly describes the basic building blocks. Our proposed scheme is described in Section III. The proof of correctness for the scheme is presented in Section IV. Section V discusses the implementation along with the performance analysis results. Section VI finally concludes our work.

\section{Preliminaries}

This section provides a brief overview on RPL, rank attacks in RPL followed by a discussion on a cryptographic technique used to secure our scheme.

\subsection{Brief Overview on RPL}

RPL is a Distance Vector (DV) based source routing protocol which provides routing support for both multicast and unicast data transfer. It creates a routing topology known as Destination Oriented Directed Acyclic Graph (DODAG) oriented towards
the root node with the help of a Directed Acyclic Graph (DAG) \cite{c1}. Each node in a DODAG is assigned a rank and is identified by a unique node ID. The nodes are organized in the form of a parent-child relationship where rank indicates the position of a node relative to other nodes in the network, respective to a DODAG root, and is determined by an Objective Function (OF). MRHOF \cite{c4} is chosen as the OF for this work which utilises Expected Transmission Count (ETX) metric to determine paths with the smallest cost.

\begin{figure} [!ht]
\begin{center}  
\includegraphics[scale=0.50]{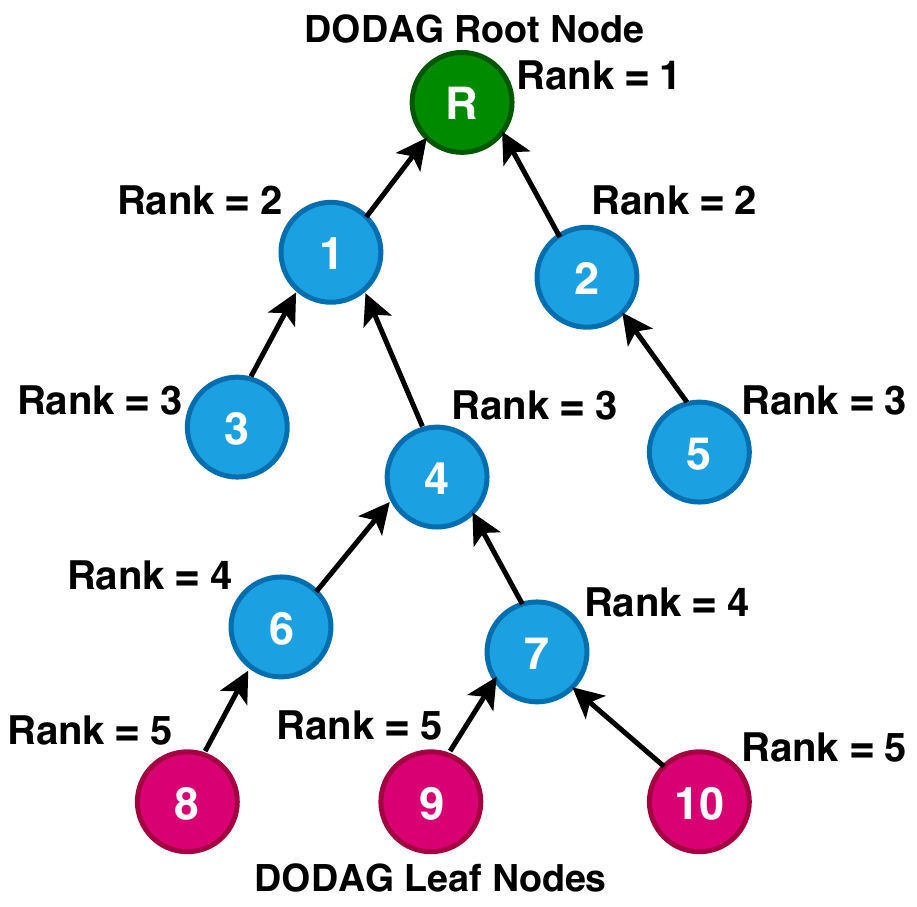}
\setlength{\belowcaptionskip}{-10pt}
\caption{\small \sl DODAG Topology \label{fig:Image2}}
\end{center}  
\end{figure}

For the construction of DODAG topology, the root initially broadcasts a DAG Information Object (DIO) control message. Upon receiving single/multiple DIO messages, a node willing to join the DODAG determines its own rank and then selects the preferred parent based on the least rank. Each node can also issue a Destination Advertisement Message (DAO) to propagate reverse root information in order to register each visited node over the particular upward route. A new node may also join an existing network by broadcasting a DODAG Solicitation Message (DIS) message in order to solicit DIO messages from its neighbors. RPL specifies two modes of operations to maintain downward routes : storing and non-storing modes \cite{c1}. In storing mode, all non-leaf nodes store routing information of its sub-DODAG whereas in non-storing mode, only the root stores routing information and is therefore susceptible to a single point of failure. For example, Fig. \ref{fig:Image2} shows a typical DODAG arrangement where R is the root node, nodes 8, 9 and 10 are the leaf nodes and the rest are intermediate nodes. Further, if a sample node is 1, it's child nodes are \{3,4\} whereas the parent node of \{3,4\} is 1.

\subsection{Rank Attacks in RPL}

In RPL, rank attack is the most destructive attack amongst all since it is capable of disrupting the entire network topology alongside producing other attacks like selective forwarding, sinkhole, wormhole, etc. There are two types of rank attacks in RPL, decreased and increased. In decreased rank attack, a malicious node illegitimately advertises a lower rank value to falsely locate itself closer to the root. As a result, many neighbouring nodes select the malicious node as their preferred parent to connect to the root \cite{c5}. On the other hand, in increased rank attack,  a malicious node voluntarily increases its rank by falsifying the DIO messages in order to disrupt the routing topology. This may lead to a scenario where a child node in the sub DODAG of the malicious node has a rank lower than its parent, thus forcing it to select a far-away node as a new preferred parent \cite{c5}.

\subsection{Attacker Model \label{2c}}

We consider an attack model where initially all the sensor nodes are legitimate. Malicious nodes join the network after the formation of DODAG topology. The malicious nodes may launch rank attack immediately after joining the network or after performing sometime as legitimate. Alternatively, there are two ways by which malicious nodes may launch rank attack. In the first case, a malicious node joins a network and immediately performs rank attack. On the other hand, in the second case, a malicious node may also join a network as a legitimate node and trigger rank attack after sometime. Further, an attack may be launched by either sending incorrect rank to the sink or to the neighbouring nodes while sending correct rank to the sink. The other option of sending correct rank to the neighbour nodes while an incorrect rank to the sink doesn’t arise practically for either of the two attacks (decreased or increased), because the aim of disrupting the sub-DODAG topology is not fulfilled in such cases. 

\subsection{HMAC using LOCHA}

HMAC \cite{c6} is a specific type of MAC obtained by using a keyed-hash function over the data along with a secret key. It is primarily used to verify message integrity as well as authenticity of the data source. In this work, LOCHA \cite{c7}, a lightweight hash function is used as the underlying hash algorithm to minimize computational overhead on low-end IoT devices. For computation, the secret key ($Key$) is broken down into two parts $Key_{inner}$ and $Key_{outer}$. HMAC over message $m$ is calculated as follows, where H is the hash function.
\vspace{-5pt}
\begin{equation}
    HMAC(m,Key) = H (Key_{outer}||H (Key_{inner}||m))
\end{equation}

\section{Proposed Rank Attack Detection Scheme}

This section describes our proposed \textbf{L}ow Overh\textbf{E}ad Rank \textbf{A}ttack \textbf{D}etection scheme for s\textbf{E}curing \textbf{R}PL based IoT (\textbf{LEADER}) applicable for non-storing operation mode of RPL. The proposed method executes a distributed module running in all the participating nodes and a centralized module in the sink to detect rank attacks. Further, HMAC-LOCHA, a lightweight Message Authentication Code (MAC) algorithm has been applied within the nodes to preserve basic security by verifying integrity and authenticity \cite{c18} of the control messages exchanged between nodes and sink. The sink is assumed to have sufficient computation and memory resources and hence cannot be compromised.

\subsection{Modification of DAO Control Message}

\begin{figure}[!ht]
\centering
\subfloat[Original Format]{%
  \includegraphics[clip,height=0.4in, width=\columnwidth]{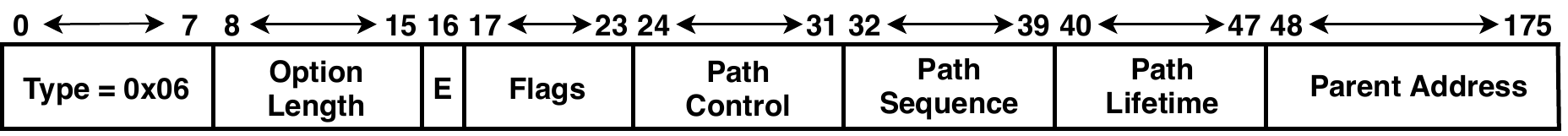}%
  \label{fig:Image1a}
}

\subfloat[Modified Format]{%
  \includegraphics[clip,height=0.4in, width=\columnwidth]{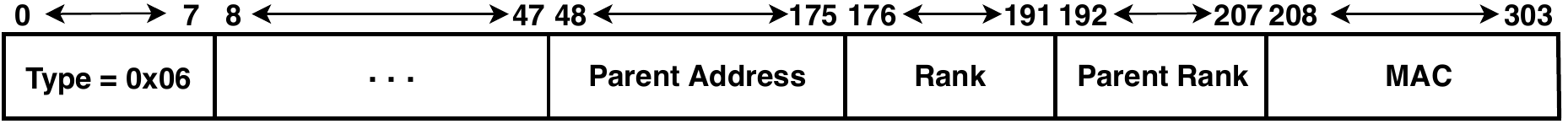}%
  \label{fig:Image1b}
}
\setlength{\belowcaptionskip}{-5pt}
\caption{\small \sl Transit Information Option of DAO control message} \label{fig:Image1}
\end{figure}

In order to execute our proposed scheme, certain additional information needs to be forwarded to the sink which are appended in the DAO control message. This information is added by each node with the Transit Information option of DAO control message shown in Fig. \ref{fig:Image1a} while sending it to the sink. Fig. \ref{fig:Image1b} shows the modified Transit Information option where the additional fields are : (i) Parent rank (16 bits): rank of the preferred parent received through DIO control messages, (ii) Rank (16 bits): rank of a node calculated using objective function specified in DIO, (iii) MAC (96 bits): generated from the additional information and the secret key [obtained using Eq. (1)]. So, 128 bits are added to the DAO control message making its size 78 bytes.

\subsection{Principle of Operation \label{3b}} 

Whenever node $N$ wants to send a DAO control message to the sink after selecting a preferred parent $P$, the distributed module on node $N$ gets triggered. The module computes a MAC over certain parameters and then adds 3-tuples including the computed MAC value to the DAO control message before forwarding it to the sink. The sink maintains an Information Table to store 4-tuple information about all nodes from the received DAO messages. Now, the centralized module running in the sink extracts the information received through DAO message from the node $N$. It then re-calculates the MAC value. If both the MAC values match, then sink is assured that the message has indeed been sent by node $N$ as well as the integrity of the received DAO message is preserved. 

Upon successful verification of the MAC values, the sink starts checking the rank of the node $N$. In case $N$ is a leaf node, then the sink checks for decreased or increased rank attack conditions by using either of the two threshold values, Minimum-Hop-Rank-Increase or Maximum- Feasible-Rank-Increase respectively. However, if $N$ is an intermediate node, primarily the sink checks the rank of $N$ received from its child node/s. If the rank doesn’t match with the rank received from DAO message, then sink declares node $N$ as malicious. Otherwise, the sink applies threshold checking similar to the leaf nodes for detecting rank attack. Thus, our scheme efficiently addresses both the scenarios mentioned in Section \ref{2c}: (i) node $N$ sends its incorrect rank to the sink (ii) node $N$ sends an incorrect rank to its neighbouring node while sending correct rank to the sink.

\subsection{Rank Attack Detection \label{3c}}

The type of rank attack is detected by checking the rank inconsistency among nodes. Rank inconsistency with respect to parent-child relationship is identified by comparing the node rank ($R_N$) with its parent node rank ($R_P$).

\begin{itemize}[leftmargin=*]
    \item \textbf{Decreased Rank Attack Detection:} The rank inconsistency checking is based on the Minimum-Hop-Rank-Increase value, defined in RFC 6550, which acts as a threshold. If the following condition holds, then the centralized module detects node $N$ to be a decreased rank attacker node.
\vspace{-5pt}
\begin{equation}
    R_N < R_P+Minimum-Hop-Rank-Increase
\end{equation}

\item \textbf{Increased Rank Attack Detection:} The rank inconsistency checking is based on Maximum-Feasible-Rank-Increase threshold whose value is not static. It depends on the current DODAG topology and is calculated for each parent node in the DODAG. To calculate this threshold value for a parent node $P$, the proposed scheme finds a child node of $P$ bearing minimum rank ($R_{child}$) amongst all its siblings. The threshold for $P$ is then calculated as:
\vspace{-5pt}
\begin{multline*}
\noindent Maximum-Feasible-Rank-Increase=(R_{child}-R_P)\\*(1+Parent-Switching-Threshold) 
\end{multline*}
\noindent where, Parent-Switching-Threshold value is 1.5.
Otherwise, if $P$ has a single child then threshold is calculated as:
\vspace{-10pt}
\begin{multline*}
\noindent
Maximum-Feasible-Rank-Increase=2*\\(1+Parent-Switching-Threshold)
\end{multline*}
\end{itemize}

\noindent Finally, if the following condition holds, then the centralized module detects $N$ as an increased rank attacker node.
\vspace{-5pt}
\begin{equation}
    R_N > R_P+Maximum-Feasible-Rank-Increase
\end{equation}

\subsection{Algorithms}

This section provides both the algorithms designed as distributed module (Algorithm \ref{algo:Algo1}) and centralized module (Algorithm \ref{algo:Algo2}).

\begin{algorithm}[!htb]
\small
    \SetKwInOut{Input}{Input}
    \SetKwInOut{Output}{Output}
	\SetKwInOut{Runs}{Runs}
	\SetKwInOut{Triggers}{Triggers}
	\Input{DAO Control Message}
	\Output{DAO message with additional information}
	\Runs{On all nodes other than sink}
	\Triggers{Before sending DAO control message to sink}
	Generate $MAC_1=HMAC((ID_N||R_N||ID_P||R_P),K)$\\
	ADD $R_N$, $R_P$, $MAC_1$ into DAO control message\\
	Send DAO to sink node
	\caption{Distributed Module \label{algo:Algo1}}
\end{algorithm}

\begin{algorithm}[!htb]
\small
\DontPrintSemicolon
    \SetKwInOut{Input}{Input}
    \SetKwInOut{Output}{Output}
	\SetKwInOut{Runs}{Runs}
	\SetKwInOut{Triggers}{Triggers}
	\Input{DAO Control Message}
	\Output{Set of Malicious Node IDs}
	\Runs{On the sink node}
	\Triggers{When sink node receives DAO control messages from other nodes in the network}
	Retrieve $ID_N$, $R_N$, $ID_P$, $R_P$ and $MAC_1$ from the received DAO control message\\
	Generate $MAC_2=HMAC((ID_N||R_N||ID_P||R_P),K)$ from the received values\\
	\If{($MAC_1$ == $MAC_2$)}{
	Insert $<ID_N$, $R_N$, $ID_P$, $R_P>$ into Information Table\\
	\If{(Node $N$ is a parent node)}{
	$R_{CP} = getChildParentRank(ID_N)$\\
	\If{($R_N\neq R_{CP}$)}{
	Insert $ID_N$ into Malicious\_Nodes \tcp{\footnotesize a set to store malicious node IDs}
	exit()
	}
	}
	$Maximum-Feasible-Rank-Increase$ = getMFRI($ID_P,R_P$)\\
	\If{(condition (2) || condition (3) holds)}{
	Insert $ID_N$ into Malicious\_Nodes
	}
	}\Else{
	Discard the DAO control message\\
	}
	\SetKwProg{Fn}{Function}{:}{}
\SetKw{KwEnd}{end}

\SetKwFunction{FCPR}{getChildParentRank}
\Fn{\FCPR{$ID_{Node}$}}{
\For{(Entry $i$ in the Information\_Table)}{
\If{($ID_{Node}$ is the parent node)}{
\KwRet {$ParentRank$}\;
}
}
}
\KwEnd
\;

\SetKwFunction{MFRI}{getMFRI}
\Fn{\MFRI{$ID_{Parent}, R_{Parent}$}}{
$R_{child}$ = Maximum\_Rank\\
\For{(Entry $i$ in the Information\_Table)}{
\If{(child node having minimum rank with parent as $ID_{Parent}$)}{
$R_{child}=NodeRank$
}
}
\If{(Node $P$ has more than one child)}{
\KwRet {$(R_{child}-R_P)*PST$} \tcp{\footnotesize PST is parent switching threshold value}
}
\KwRet {$2*(1+PST)$}\;
}
\KwEnd
\;

	\caption{Centralized Module \label{algo:Algo2}}
\end{algorithm}

\noindent \textbf{Complexity Analysis:} Algorithm \ref{algo:Algo1} consists of two tasks. Generating $MAC_1$ value which takes O(HMAC) time, where O(HMAC) is the complexity of the HMAC algorithm. Adding information to the DAO control message and sending it to the sink is done in constant time. So, overall time complexity of this algorithm is O(HMAC) + O(1) = O(HMAC). Similarly, Algorithm \ref{algo:Algo2} first generates $MAC_2$ and compares it with $MAC_1$ which again takes O(HMAC) time. For rank inconsistency checking, sink uses two functions each of which takes O(n) time, where n is the number of nodes in the network. All other operations take constant time. So, overall time complexity of Algorithm \ref{algo:Algo2} is O(HMAC) + O(n) + O(n) + O(1) = O(n).

\section{Proof of Correctness}

This section establishes the correctness of the proposed scheme for both decreased and increased rank attacks. 

\begin{figure}[!ht]
\centering
\subfloat[Decreased Rank Attack]{%
  \includegraphics[clip,height=1.8in, width=0.50\columnwidth]{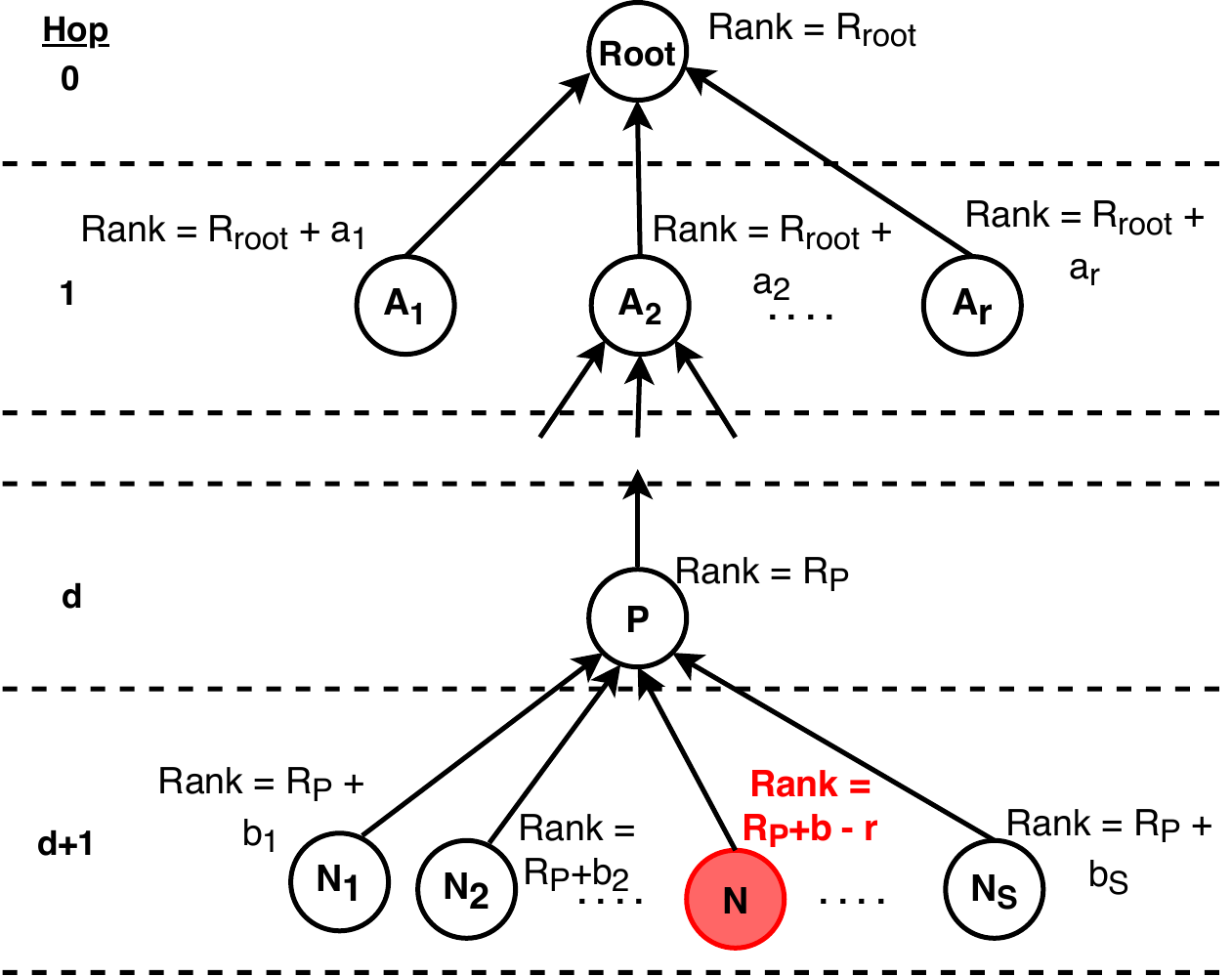}%
  \label{fig:Image4a}
}
\subfloat[Increased Rank Attack]{%
  \includegraphics[clip,height=1.8in, width=0.50\columnwidth]{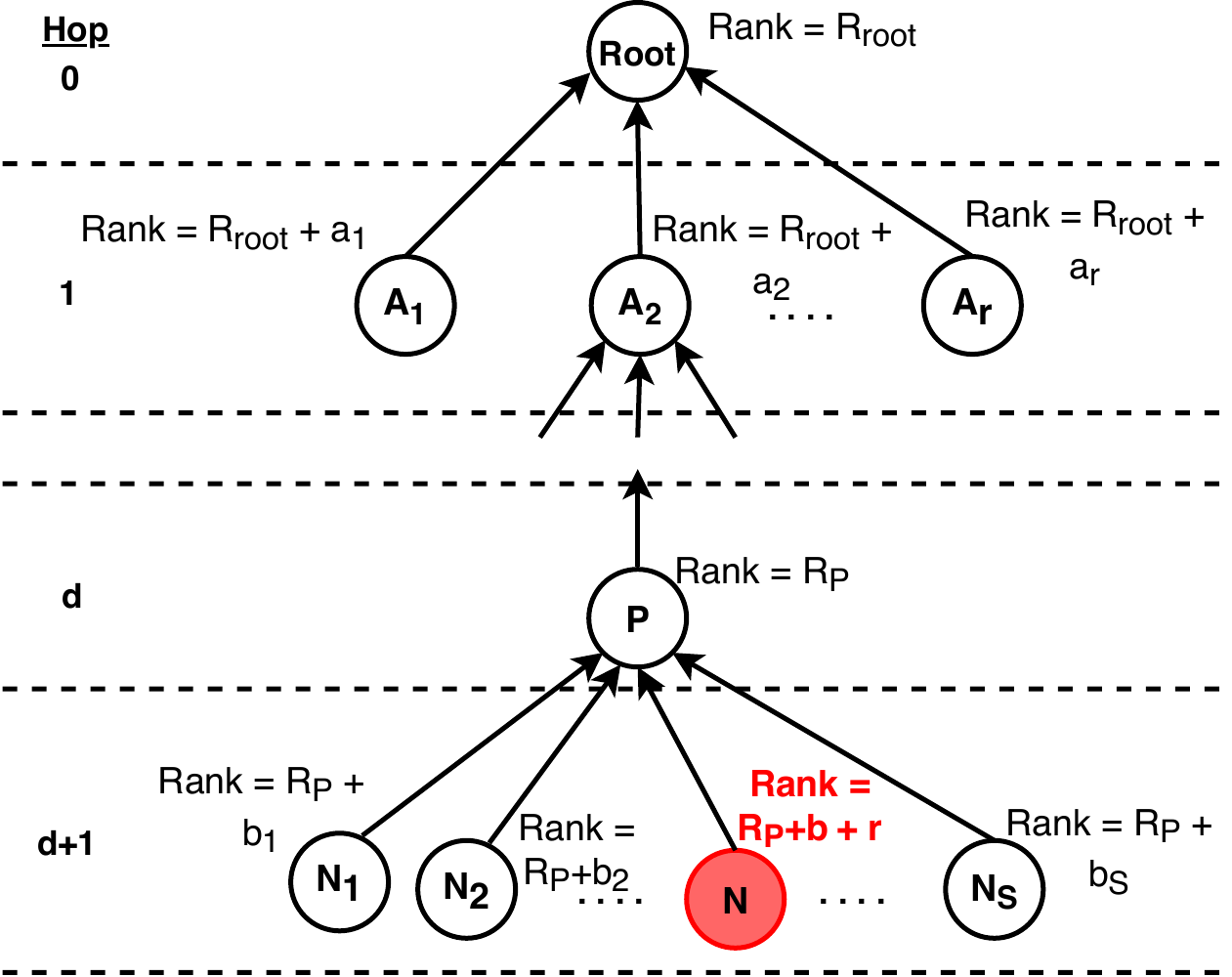}%
  \label{fig:Image4b}
}
\setlength{\belowcaptionskip}{-5pt}
\caption{\small \sl A RPL Network in presence of malicious nodes} \label{fig:Image4}
\end{figure}

\noindent\textbf{Theorem:} \textit{If a RPL DODAG contains $n >= 2$ nodes and the nodes are distributed within $D >= 1$ hop distance from the root node, then the proposed scheme is able to detect malicious nodes placed at any hop distance $d$ from the root where range of $d$ is between 1 and $D$.}

\noindent Let us consider a RPL network where node $P$ is placed at $d$ hop distance from the root and rank of node $P$ is $R_P$. The $s$ number of nodes, $N_1$, $N_2$ ... $N_s$ are placed at hop distance $d+1$ from the root that chose $P$ as their preferred parent. Ranks of nodes $N_1$, $N_2$ ... $N_s$ are $(R_P+b_1)$, $(R_P+b_2)$, ..., $(R_P+b_s)$ respectively, where $b_1$, $b_2$, ..., $b_s$ are the ETX values between respective nodes and their preferred parent $P$.

\noindent \textbf{Proof:} A malicious node $N$ may either join a network and falsely changes its rank from $(R_P + b)$ to $(R_P + b - r)$ or it may join as a legitimate node and later decreases its rank to $(R_P + b - r)$ where $r$ is a positive integer. In both these cases, as shown in Fig. \ref{fig:Image4a}, the updated false information of the node will be sent to root through DAO message. The root checks condition (2) of Section \ref{3c} and if it is satisfied then declares $N$ as a decreased rank attacker node. Similarly, $N$ may increase its rank from $(R_P + b)$ to $(R_P + b + r)$ which is shown in Fig. \ref{fig:Image4b}. Here, the root checks condition (3) of Section \ref{3c} and declares $N$ as an increased rank attacker node provided the condition is true.

The above theorem can then be extended to formulate a corollary to show the number of malicious nodes that our scheme can tolerate.

\subsection{Malicious Node/s Tolerance along a Specific Path from Root to a Leaf}

\noindent \textbf{Corollary:} \textit{If a path constitutes of $n$ consecutive sequence of nodes from root to a specific leaf, then our proposed scheme can tolerate at least one and at most $n-1$ (except root) malicious nodes in that path.}

\begin{itemize}[leftmargin=*]
    \item \textit{Best Case Scenario:} Let us consider a single node to be malicious along a path from root to leaf. If $N$ is a decreased or increased rank attacker node, then the detection module of root checks condition (2) or (3) and declares it as malicious. 
    \item \textit{Worst Case Scenario:} Let us consider all nodes except root to be malicious along a path from root to a specific leaf. If these nodes perform decreased or increased rank attack then initially, root finds its first hop child node to be malicious using the condition (2) or (3) and adds it to the malicious nodes list. Eventually, root checks rank inconsistency of nodes belonging to successive hops which also turn out to be malicious and is added to the same list. 
\end{itemize}

\subsection{Malicious Node/s Tolerance in a Network}

Let us consider a RPL DODAG as a m-ary complete tree where leaf nodes are placed at $D$ hop distance from the root. Therefore, total number of nodes $n$ in the network = $1 + m + m^2 + \ldots + m^D = \frac{m^{D+1}-1}{m-1}$\\

\noindent \textbf{Decreased Rank Attack:} Minimum number of legitimate nodes required is 1 (Root only). So, maximum number of malicious nodes tolerated = $\frac{m^{D+1}-1}{m-1}-1$ = $(n-1)$ nodes

\noindent \textbf{Increased Rank Attack:} The proposed scheme assumes that at least one child node of each parent should be legitimate. Thus, minimum number of legitimate nodes required = 1 (root) + $(1 + m + m^2 + \ldots + m^{D-1})$ = $(1+\frac{m^D-1}{m-1})$\\
Number of malicious nodes tolerated atmost = $\frac{m^{D+1}-1}{m-1}-(1+\frac{m^D-1}{m-1})$

\section{Performance Evaluation}

Performance of the LEADER is evaluated and compared with SBIDS \cite{r4} both theoretically and experimentally.

\subsection{Theoretical Analysis}

This section evaluates the performance of both LEADER and SBIDS theoretically in terms of storage, communication and computation overheads. We consider a network consisting of a sink node and $n$ number of sensor nodes. The analysis is done based on sky mote \cite{c1,m1} specifications. Here, we primarily focus on overheads incurred by the control messages, especially DAO control messages.

\subsubsection{Storage Overhead}

In both the schemes, sink uses an additional Information Table for storing information sent along with the DAO control messages.

\noindent \textbf{LEADER:} The sink stores $n$ entries for $n$ nodes in the Information Table where each tuple is of size 2 bytes. The sink also stores a 16-byte secret key and two substitution tables to generate LOCHA based MAC which requires to store $(97+68) = 165$ integers, i.e. $(165*2) = 330$ bytes. All other nodes store only the secret key and 330 bytes for two substitution tables. Thus, total storage overhead:
$$S_{total}={(8n+16+330)+ n(16+330)}= (354n+346)\ bytes$$

\noindent \textbf{SBIDS:} SBIDS stores one extra tuple in the Information Table. Further, the sink also stores 16-bytes secret key and two substitution tables for using AES-128 which requires $(256 + 256) = 512$ bytes. All other nodes store only 16-bytes secret key and 512 bytes for two substitution tables. Therefore, the total storage overhead:
$$S_{total}={(10n+16+512)+ n(16+512)} = (538n+528)\ bytes$$

\subsubsection{Communication Overhead}

The difference in communication overhead between the two schemes lies on the DAO control message size. So, here we consider overheads due to a single DAO control message communication.

\noindent \textbf{LEADER:} Let us assume a node deployed at $d$ hop distance from sink sends a DAO message (78 bytes) to the sink through  $(d-1)$ intermediate nodes. To start with, the node sends the message to its parent, which incurs 78 bytes overhead. Next, each of the intermediate nodes receiving the DAO message forwards it towards the sink. Thus, communication overhead of one intermediate node is $(78 + 78)$ bytes = $156$ bytes. At last, sink receives the DAO message of 78 bytes. According to the specification of Tmote Sky \cite{c1,m1}, energy consumption for transmitting and receiving 1-byte of data is 0.00189 mJ and 0.00167 mJ respectively. Thus, the communication overhead in terms of energy  ($E_{comm}$) can be expressed as follows:
\begin{dmath*}
E_{comm}=[(78+(d-1)*78)*0.00189 ]+[(((d-1)*78)+78)*0.00167]\ mJ
\end{dmath*}
$$=[(d*78)*0.00189]+[(d*78)*0.00167] = 0.277d\ mJ$$

\noindent \textbf{SBIDS:} In SBIDS, the DAO message size is 80 bytes. Thus, the communication overhead in terms of energy ($E_{comm}$) can be expressed as follows:
\begin{dmath*}
E_{comm}=[(80+(d-1)*80)*0.00189 ]+[(((d-1)*80)+80)*0.00167]\ mJ
\end{dmath*}
$$=[(d*80)*0.00189]+[(d*80)*0.00167]=0.285d\ mJ$$

\subsubsection{Computation Overhead}

We consider computation overhead incurred due to additional operations performed by the nodes other than normal RPL operations. Table \ref{tab:table1} provides number of computation cycles required to implement different relevant algorithmic constructs for both the competing schemes including ours.

\noindent \textbf{LEADER:} Computation overhead is incurred by sink and DAO sender node for the following tasks:\\
Sender node: MAC generation using HMAC-LOCHA\\
Sink: MAC verification (generation + comparison) + Rank attack detection by consulting Information Table

\begin{savenotes}
\begin{table}[!ht]
\centering
\caption{Required number of cycles for relevant algorithmic constructs \cite{m2,c10,c11}}
\label{tab:table1}
\resizebox{\columnwidth}{!}{%
\begin{tabular}{c|c}
\hline
\textbf{Algorithmic Constructs} & \textbf{No. of Cycles} \\ \hline
ADD/SUBTRACT, MOV, XOR & 4 \\ \hline
SEARCH & \begin{tabular}[c]{@{}c@{}}(2 * BR + 5 * MOV + INX + DCR +\\ CMP + 2 * JNZ + 2 * STA + 2 * HLT)\\= (2x3 + 5x4 + 5 + 4 + 6 + 2x3 + 2x5 + 2x6) = 69\end{tabular} \\ \hline
IF-ELSE & (JMP + JL + CMP) = (2 + 2 + 6)   = 10 \\ \hline
HMAC-LOCHA & 6032 \\ \hline
AES-128 Encryption, Decryption & \begin{tabular}[c]{@{}c@{}}151280, 192480\footnote[2]{Cycles required per byte for Encryption/Decryption is 1891/2406} \end{tabular} \\ \hline
\end{tabular}%
}
\end{table}
\end{savenotes}

\noindent Now, consulting Table \ref{tab:table1}, the details of computation overhead is provided in Table \ref{tab:table2}. According to the Sky mote specification, the clock of MSP430 microcontroller is up to 8 MHz \cite{c1,m1}. So, the energy needed to run one cycle = $3*1.8*0.125=0.675nJ$. Thus, the computation overhead in terms of energy ($E_{comp}$) can be expressed as follows:
$$E_{comp}=(6032+6298)*0.675\ nJ=8323\ nJ$$

\noindent \textbf{SBIDS:} Computation overhead incurred by SBIDS is as follows:\\
Sender node: Encryption of DAO control message\\
Sink: Decryption of DAO control message + Rank attack detection consulting Information Table

\noindent Now, again consulting Table \ref{tab:table1}, the details of computation overhead is provided in Table \ref{tab:table2}. Thus, similar to LEADER the computation overhead in terms of energy ($E_{comp}$) for SBIDS can be expressed as follows:
$$E_{comp}=(151280+192785)*0.675\ nJ=232243\ nJ$$

\begin{table}[!ht]
\centering
\caption{Required additional number of cycles for a single DAO communication}
\label{tab:table2}
\resizebox{\columnwidth}{!}{%
\begin{tabular}{c|c|c|c}
\hline
\multicolumn{2}{c|}{\textbf{Method}} & \textbf{Algorithmic Construct} & \textbf{No. of  Cycles} \\ \hline
\multirow{2}{*}{LEADER} & Sender & HMAC-LOCHA & 6032 \\ \cline{2-4} 
 & Sink & \begin{tabular}[c]{@{}c@{}}HMAC-LOCHA, 8*IF-ELSE,\\ 9*MOV, 3*ADD, 2*SEARCH\end{tabular} & \begin{tabular}[c]{@{}c@{}}6032 + 8*10 + 9*4 \\ +3*4 + 2*69 = 6298\end{tabular} \\ \hline
\multirow{2}{*}{SBIDS} & Sender & Encryption (AES-128) & 151280 \\ \cline{2-4} 
 & Sink & \begin{tabular}[c]{@{}c@{}}Decryption (AES-128), 7*IF-ELSE,\\ 4*MOV, 3*ADD, 3*SEARCH\end{tabular} & \begin{tabular}[c]{@{}c@{}}192480 + 7*10 + 4*4\\ + 3*4 + 3*69 = 192785\end{tabular} \\ \hline
\end{tabular}%
}
\end{table}

\subsubsection{Comprehensive Evaluation} Finally, we provide comprehensive overhead comparison of LEADER and SBIDS in Table \ref{tab:table3}. We evaluate the theoretical overheads by considering number of sensor nodes $n$ = 50 and hop distance of a node $d$ = 5 for both the schemes. We observe, from Table \ref{tab:table3}, theoretical overheads for LEADER are considerably less than that of SBIDS in terms of all the three overheads. So, we can easily claim that the proposed scheme is low overhead compared to the competitor SBIDS.

\begin{table}[!ht]
\centering
\caption{Comprehensive overhead comparison}
\label{tab:table3}
\resizebox{\columnwidth}{!}{%
\begin{tabular}{c|c|c|c}
\hline
\textbf{Method} & \textbf{\begin{tabular}[c]{@{}c@{}}Storage\\ Overhead (bytes)\end{tabular}} & \textbf{\begin{tabular}[c]{@{}c@{}}Communication \\ Overhead (mJ)\end{tabular}} & \textbf{\begin{tabular}[c]{@{}c@{}}Computation\\ Overhead (nJ)\end{tabular}} \\ \hline
LEADER & \begin{tabular}[c]{@{}c@{}}354 * N + 346 = \\ 354 * 50 + 346 = 18046\end{tabular} & \begin{tabular}[c]{@{}c@{}}d * 0.277 = \\ 5 * 0.277 = 1.38\end{tabular} & 8323 \\ \hline
SBIDS & \begin{tabular}[c]{@{}c@{}}538 * N + 528 = \\ 538 * 50 + 528 = 27428\end{tabular} & \begin{tabular}[c]{@{}c@{}}d * 0.285 = \\ 5 * 0.285 = 1.43\end{tabular} & 232243 \\ \hline
\end{tabular}%
}
\end{table}

\subsection{Experimental Evaluation}

This section compares the performance of LEADER with the state-of-the-art competitor SBIDS in terms of several metrics through simulation.

\subsubsection{Simulation Environment} We use Cooja Simulator based on well-known Contiki operating system \cite{c13}. Similar to theoretical analysis, in simulation also we use Tmote sky mote \cite{c1,m1}, a well-known IoT sensor device. We use a power profiling tool provided by Contiki called PowerTrace \cite{c15} to compute the network energy consumption. The average network power consumption is computed based on the nominal values of current consumption and supply voltage \cite{c13,c15} of Tmote Sky sensor.

\noindent The rank attacks have been implemented based on the RPL library within Contiki operating system. We consider that attacker nodes send decreased ranks through DIO message to the neighbour nodes. We simulate a static network with 50 nodes, deployed randomly, including the sink. An average result of 10 independent runs is taken while plotting the simulation graphs. Table \ref{tab:table4} summarizes the other important simulation parameters used here.

\begin{table}[!ht]
\centering
\caption{Simulation Parameters}
\label{tab:table4}
\resizebox{0.85\columnwidth}{!}{%
\begin{tabular}{c|c}
\hline
\textbf{Parameter Name} & \textbf{Values} \\ \hline
Simulation Area & 300$\times$300 m \\ \hline
Radio Medium & Unit Disk Graph Radio Medium (UDGM) \\ \hline
Mote Startup Delay & 1000 ms \\ \hline
MAC Layer & IEEE 802.15.4 \\ \hline
Radio Model & CC2420 \\ \hline
Transmission Range & 50 m \\ \hline
Objective Function & MRHOF \\ \hline
Packet Sent Interval & 2m \\ \hline
\end{tabular}%
}
\end{table}

\subsubsection{Simulation Metrics} 

The performance evaluation of the proposed scheme is primarily divided into two stages: measuring extent of achieving major design goals and evaluation of network performance. The design goals are assessed by design parameters such as rank attack detection accuracy rate, false positive rate, false negative rate and total energy consumption. On the other hand, network performance metrics are detection latency (or delay) and packet delivery ratio. 

\noindent \textbf{Accuracy Rate:} The detection accuracy is defined as the ratio of the number of nodes that are detected correctly (legitimate or attacker) to the total number of nodes present in the network \cite{c14}.

\noindent \textbf{False Positive Rate (FPR):} It is defined as the ratio of the number of legitimate nodes that are incorrectly detected as attacker nodes to the total number of legitimate nodes present in the network \cite{c14}.

\noindent \textbf{False Negative Rate (FNR):} It is defined as the ratio of the number of attacker nodes that are incorrectly detected as legitimate nodes to the total number of attacker nodes present in the network \cite{c14}.

\noindent \textbf{Total Energy Consumption:} It is defined as the average energy consumed by the nodes in the network over a time period \cite{c9}. 

\noindent \textbf{Latency:} Latency (or delay) in identification of malicious node is defined as the difference between the time at which a malicious node joins the network to the time when sink detects the node as malicious \cite{r4}. 

\noindent \textbf{Packet Delivery Ratio (PDR):} It is defined as the ratio of the number of packets delivered to the destination (sink) to the total number of packets sent to the destination (sink) by all the sender nodes over a period of time \cite{c1}.

\subsubsection{Results and discussion} We consider simulation results of both the schemes LEADER and the SBIDS in without-basic-security and with-basic-security modes. This is done to measure the additional overheads incurred by each of the schemes for incorporating security features. In maintaining such securities, the LEADER uses a cryptographic MAC (HMAC using LOCHA) and the SBIDS uses cryptographic encryption and decryption (AES-128).

\begin{figure} [!ht]
\begin{center}  
\includegraphics[scale=0.18]{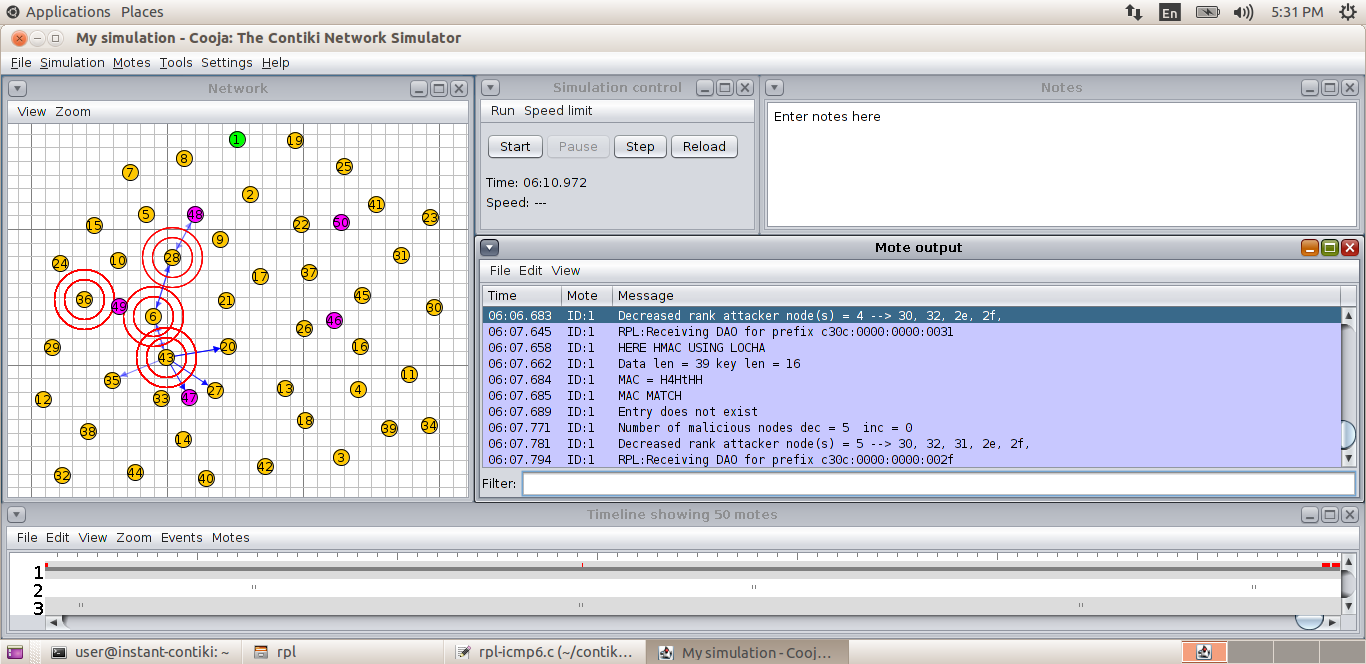}
\setlength{\belowcaptionskip}{-5pt}
\caption{\small \sl Snapshot of simulation in Cooja Simulator \label{fig:Image7}}
\end{center}  
\end{figure}

\noindent Fig. \ref{fig:Image7} shows a snapshot of the simulation carried out in Cooja with 10\% malicious nodes (i.e. 5 nodes) where green node is the sink and yellow nodes are the legitimate ones. The malicious nodes are represented in pink and located randomly in the network.

\noindent \textbf{Achieving Design Goals}

\noindent We conduct three sets of experiments to measure the extent at which design objectives have been attained for both the schemes. 

\begin{figure} [!ht]
\begin{center}  
\includegraphics[scale=0.35]{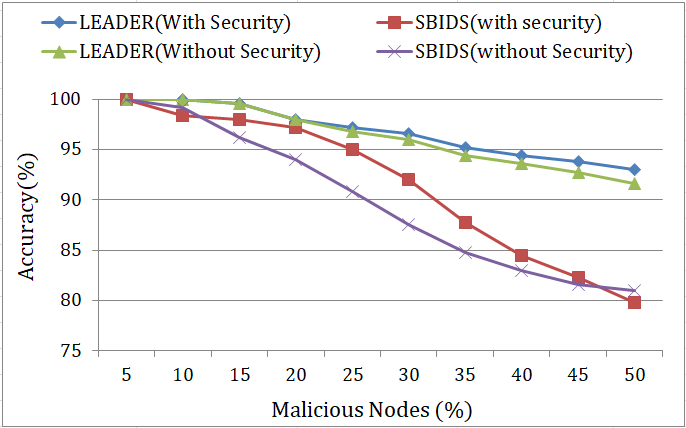}
\setlength{\belowcaptionskip}{-5pt}
\caption{\small \sl Accuracy Rate in Decreased Rank Attack Detection \label{fig:Image8}}
\end{center}  
\end{figure}

In the first set of experiment, we plot (Fig. \ref{fig:Image8}) accuracy rate of malicious node detection with varying number of malicious nodes in the network. We observe that accuracy rate decreases with increasing number of malicious nodes in both without-security and with-security cases for both the schemes. The reason for such results is sometimes a part of the network gets disconnected from the rest of the network due to rank attacks. In the network partitioning scenario, malicious nodes present in one part may not be detected by the sink as DAO messages from such nodes do not reach the sink. However, this decreasing rate of accuracy for LEADER is very less compared to SBIDS. For example, considering 50\% malicious nodes, the accuracy rate of LEADER is 13.1\% higher than SBIDS in without-security mode whereas it is 16.5\% higher in with-security mode. This is because SBIDS checks rank inconsistency for a particular node using only the received DAO message. On the contrary, as mentioned in Section \ref{3b}, LEADER performs an additional rank inconsistency checking by retrieving malicious node’s rank from its child node/s as well. This result in considerable better performance of LEADER compared to SBIDS.

\begin{figure} [!ht]
\begin{center}  
\includegraphics[scale=0.32]{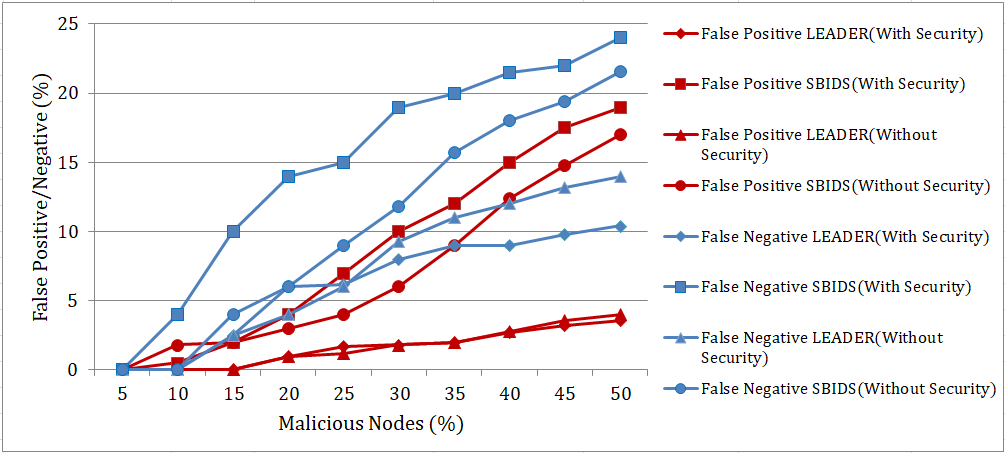}
\setlength{\belowcaptionskip}{-10pt}
\caption{\small \sl False Positive and Negative Rates in Decreased Rank Attack Detection \label{fig:Image9}}
\end{center}  
\end{figure}

In the second set of experiment, we plot (Fig. \ref{fig:Image9}) false positive and false negative rates in rank attack detection with varying number of malicious nodes. We observe that both the rates increase with increasing number of malicious nodes in the network for both the schemes in without-security and with-security cases. Similar to the accuracy rate, here also due to occasional network partitioning, the false positive/negative rates increase with increasing number of malicious nodes. However, this increase in rates is much lower in LEADER compared to the SBIDS. Precisely, the false positive rate for LEADER is lower by 71\% and 76\% whereas false negative rate for LEADER is lower by 25.6\% and 57\% on average for without-security and with-security respectively than SBIDS. Here also the reason for LEADER’s better performance is similar to the first set of experimnent.

\begin{figure} [!ht]
\begin{center}  
\includegraphics[scale=0.30]{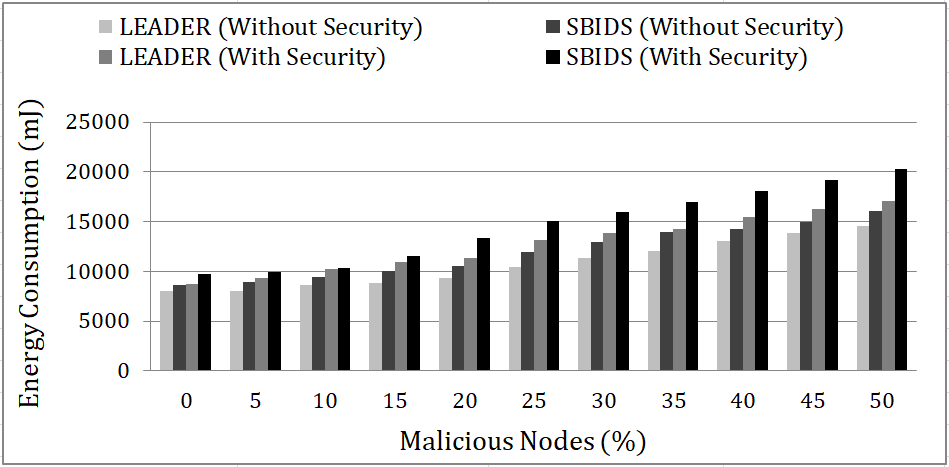}
\setlength{\belowcaptionskip}{-10pt}
\caption{\small \sl Total Energy Consumption in Decreased Rank Attack Detection} \label{fig:Image10}
\end{center}
\end{figure}

In the third set of experiment, we plot (Fig. \ref{fig:Image10}) total energy consumption with varying number of malicious nodes. It is observed that for both LEADER and SBIDS, energy consumption increases with increasing number of malicious nodes in both without-security and with-security cases. This is because increasing number of malicious nodes causes exchange of more control messages (DIO and DAO messages) among the nodes. Precisely, in presence of malicious nodes, energy consumption for LEADER is 9\%-13\% lower in without-security and 5\%-16\% lower in with-security as compared to SBIDS. Also, in absence of malicious nodes, energy consumption for LEADER is on an average 6.9\% lower in without-security and 10.5\% lower in with-security as compared to SBIDS. The reason is in SBIDS, the energy consumption increases due to the exchange of extra DAO messages. Moreover, in with-security scenario SBIDS consumes more energy than LEADER as AES-128 is more intensive compared to HMAC using LOCHA. 

\noindent \textbf{Evaluation of Network Performance}

We conduct two sets of experiments to observe network performance while achieving the design goals.

\begin{figure} [!ht]
\begin{center}  
\includegraphics[scale=0.32]{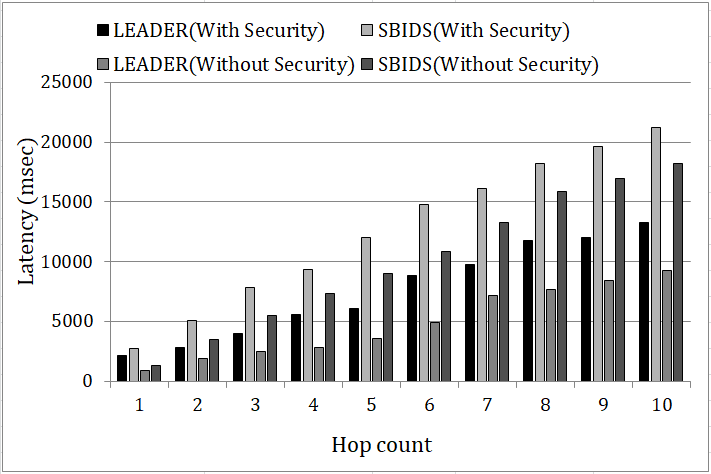}
\setlength{\belowcaptionskip}{-10pt}
\caption{\small \sl Detection Latency  in Decreased Rank Attack Detection \label{fig:Image11}}
\end{center}  
\end{figure}

In the first set of experiment, we plot (Fig. \ref{fig:Image11}) latency (or delay) of rank attack detection in presence of malicious nodes deployed at varying hop distance from the sink. For the purpose of experimentation, malicious nodes are deployed at most ten hop distances from the sink. It is observed that detection latency increases with increasing hop distance of the malicious node from the sink. This is because as hop distance increases DAO messages take more time to reach the sink which eventually takes more time in stabilizing the network. It is also observed that the detection latency of LEADER is lower by 32\%-61\% and by 22\%-49\% for without-security and with-security respectively as compared to SBIDS.

\begin{figure} [!ht]
\begin{center}  
\includegraphics[scale=0.32]{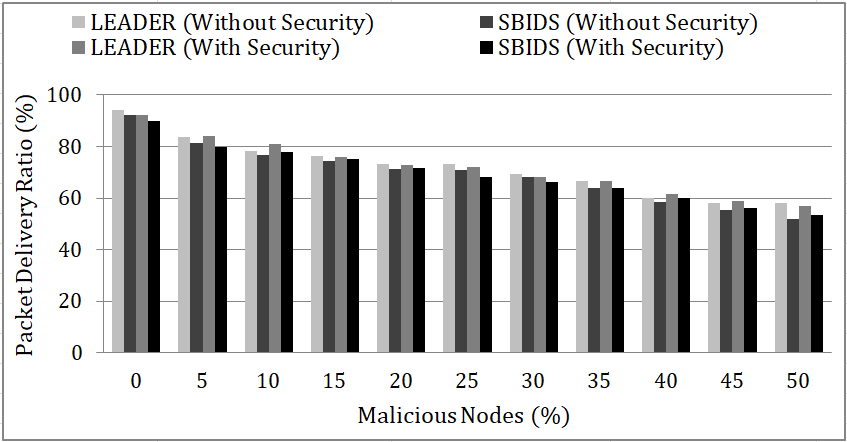}
\setlength{\belowcaptionskip}{-5pt}
\caption{\small \sl Packet Delivery Ratio in Decreased Rank Attack Detection} \label{fig:Image12}
\end{center}
\end{figure}

In the second set of experiment, we plot (Fig. \ref{fig:Image12}) packet delivery ratio with varying number of malicious nodes. It is observed from the plot that for both the schemes packet delivery ratio decreases with increasing number of malicious nodes in both without-security and with-security modes. This is because due to rank attacks launched by malicious nodes at times, a sub-DODAG may be disconnected from the DODAG. So, packets sent by nodes of the sub-DODAG do not reach the sink. As less number of packets reaches the sink, packet delivery ratio decreases with increasing number of malicious nodes. We also observe that packet delivery ratio of LEADER is marginally higher by 3.8\% and by 3.84\% on average in without-security and with-security respectively as compared to SBIDS in the presence of malicious nodes. Also, in the absence of malicious nodes, packet delivery ratio for LEADER is higher by 1.8\% in without-security and by 3\% in with-security as compared to SBIDS.

Summarily, the LEADER shows better performance over SBIDS in achieving all the design goals including accuracy rate, false positive rate, false negative rate and energy consumption while also keeping better network performance in terms of detection latency and maintaining packet delivery ratio at par with SBIDS.

\section{Conclusion}

In view of the importance of rank attacks in RPL, we propose a low-overhead rank attack detection scheme LEADER which guarantees to detect both decreased and increased rank attacks in a RPL-based IoT network while still preserving basic security during control message exchange. We modify the existing RPL algorithm by adding rank attack detection mechanism and incorporating basic security mechanism. The distributed module of the algorithm running in the participating nodes is responsible for modifying the DAO control message with some additional information. This helps the centralized module running in the sink to efficiently detect rank attacks in the network. This simple modification of the DAO message along with the use of a low-overhead MAC algorithm makes the solution low-overhead. The proof of correctness for LEADER has suitably established our claim of detecting rank attacks under certain adversarial scenarios. The performance of the proposed scheme is compared with a state-of-the-art rank attack detection scheme SBIDS both theoretically and experimentally, which showed that LEADER has achieved superior performance over SBIDS for both with and without-basic security scenarios. Simulation results show dominance of our scheme over SBIDS scheme with respect to both set of parameters including parameters for achieving design goal and network performance metrics. As a future extension, the attack detection scheme can be modified to act in storing mode of RPL as well. The scheme may be made more realistic by considering mobility of the nodes.

\bibliographystyle{unsrt}
{\footnotesize
\bibliography{references}}

\end{document}